\documentclass [11pt,a4paper] {article}

\usepackage[T1]{fontenc}
\usepackage{lmodern}

\usepackage[cp1252]{inputenc}
\usepackage{amssymb}
\usepackage{amsmath}
\usepackage{amsfonts,amssymb}
\usepackage[dvips]{graphicx}
\usepackage{bbm}
\usepackage{enumerate}

\usepackage[shortlabels]{enumitem}

\usepackage{amsthm}
\usepackage{cancel}

\usepackage{centernot}
\usepackage{mathrsfs}

\usepackage{wasysym}

\DeclareMathAlphabet{\mathpzc}{OT1}{pzc}{m}{it}

\begin{document}

\title{The holographic principle comes from finiteness of the universe's geometry}

\author{Arkady Bolotin\footnote{$Email: arkadyv@bgu.ac.il$\vspace{5pt}} \\ \emph{Ben-Gurion University of the Negev, Beersheba (Israel)}}

\maketitle

\begin{abstract}\noindent Discovered as an apparent pattern, a universal relation between geometry and information called the holographic principle has yet to be explained. This relation is unfolded in the present paper. As it is demonstrated there, the origin of the holographic principle lies in the fact that a geometry of physical space has only a finite number of points. Furthermore, it is shown that the puzzlement of the holographic principle can be explained by a magnification of grid cells used to discretize geometrical magnitudes such as areas and volumes into sets of points. To wit, when grid cells of the Planck scale are projected from the surface of the observable universe into its interior, they become enlarged. For that reason, the space inside the observable universe is described by the set of points whose cardinality is equal to the number of points that constitute the universe's surface.\bigskip\bigskip

\bigskip\bigskip
\end{abstract}

\section{Introduction}  

\noindent The holographic principle has arisen in physics as an universal relation between geometry and information both uncontradicted and unexplained by existing theories. The basis of this relation is that a spatial volume $V$ with a boundary of area $A$ is fully described by no more than $A/4\ell_{P}^2$ bits of information, i.e., 1 bit per each 4 Planck areas $\ell_{P}^2$ (where $\ell_{P}$ is the Planck scale approximately equal to $1.6 \times 10^{-35} \text{ m}$) \cite{Hooft, Susskind, Bigatti}.\bigskip

\noindent The discovered relation is not trivial. To be sure, discretizing a flat 3-space into an array of primitive cubes of edge length $\ell_{P}$ and assuming that there is a quantum harmonic oscillator in each Planck cube, one can envision a region of volume $V$ in this space as a lattice of $V/\ell_{P}^3$ oscillators. The lattice has $N^{V/\ell_{P}^3}$ states, where $N$ is the number of states of each oscillator. Provided $N \ge 2$ and $V$ is finite, one would expect the number of degrees of freedom contained in the region to grow with $V$ in the way that $V/\ell_{P}^3 \cdot \log{N}$. But, as it has been proven to hold in a wide range of examples \cite{Bousso}, the said number is not exceeding the value $A/4\ell_{P}^2 \cdot \log{N}$, where $A$ is the area of the boundary to the region. This is deeply puzzling. For example, let $V$ be the volume of a cube with the edge length $a \gg \ell_{P}$. Then, in accordance with the discovered relation, one would have $a^{3}/\ell_{P}^3 < 3a^{2}/2\ell_{P}^2$, i.e., a contradiction: $a < 3\ell_{P}/2$.\bigskip

\noindent The universality of the aforementioned relation suggests that it is an imprint of some general law in science. Usually, this law is considered to be due to string theory \cite{Kaku}. Particularly, it was observed that string theory admits a lower-dimensional description in which gravity emerges from it \cite{Ammon}. This observation was the starting point for AdS/CFT correspondence asserting that the boundary of anti-de Sitter space (AdS for short), which is used in string-based theories of quantum gravity, can be regarded as the ``spacetime'' for a conformal field theory (abbreviated as CFT), which is used to describe elementary particles \cite{Maldacena}.\bigskip

\noindent As influential as AdS/CFT has proven to be, there is growing skepticism about whether it is adequate to faithfully represent real-world systems. Suffice it to say that the spacetime on which string-based gravitational theories live has more than 4 dimensions. In that light, AdS/CFT correspondence (or at least the most famous version of it) does not provide a realistic description of gravity. As to the versions of AdS/CFT that may offer a (slightly more) realistic description of gravity, they all imply models of spacetime that are characterized by supersymmetry having no place in our universe \cite{McLerran}.\bigskip

\noindent In contrast, the present paper will demonstrate that the relation between geometry and information is due to finiteness of geometry that describes physical space.\bigskip

\section{Physical space has a finite geometry}  

\noindent There are reasonable grounds to believe that space contains a finite number of points, the corollary being that a geometry of physical space is finite \cite{Bolotin}. Let us review those grounds.\bigskip

\noindent In axiomatic set theories, it is imposed that axioms for logic and mathematics must be formulated only on pure or hereditary sets, i.e., ones endowed with no features at all. Contrastively, a vector space -- a set with linear operations defined upon it -- is not pure: It contains urelements, i.e., objects (vectors) that are not sets but may be elements of sets \cite{Weisstein}. So, in accordance with the aforesaid imposition, axioms that are new to or different from $\texttt{ZF}$, the set of the axioms of Zermelo--Fraenkel set theory, cannot be acquired by interchanging ``sets'' with ``vector spaces'' in $\texttt{ZF}$.\bigskip

\noindent But on the other hand, there are no mathematical grounds for preferring pure sets to sets containing urelements \cite{Taylor}. What is even more crucial, the metamathematical imposition of pure sets brings about one of the most serious problems in modern physics, viz., the emergence of infinities in well-formed formulas of the classical and quantum formalisms.\bigskip

\noindent Thus, it makes sense to present Hilbert space theory, fundamental to quantum mechanics, in the form of the axiomatic set theory $\texttt{ST}_{\texttt{Hil}}$ wherein ``sets'' are replaced with ``vector spaces''. As it turned out, all but two axioms of $\texttt{ZF}$ -- the axiom of empty set denoted $\textbf{\texttt{Empty}}$ and the axiom of infinity denoted $\textbf{\texttt{Inf}}$ -- can be translated into the formal language of Hilbert space theory in this manner.\bigskip

\noindent The cause for the exclusion of the axiom $\textbf{\texttt{Empty}}$ is the fact that a set of vectors must include at least the zero vector 0 to be a vector space. Therefore, the notion of empty set is not translatable into the notion of empty vector space. Furthermore, unlike the original axiom $\textbf{\texttt{Inf}}$ stating that the cardinality of a set capable of including some set $X$ together with \emph{all} successors of $X$ is \emph{infinite}, the translation of \textbf{\texttt{Inf}} into the language of Hilbert space theory asserts that the cardinality of a vector space capable of including some vector space $\mathcal{X}$ plus \emph{all} successors of $\mathcal{X}$ can be \emph{finite}, in particular 1. Indeed, any vector space must have at least one element, namely, the zero vector space $\{0\}$. Since each successor of $\{0\}$ is $\{0\}^{+} \!\mathrel{\mathop:}=\! \{0\} \cup \{\{0\}\} = \{0\} \cup \{0\} = \{0\}$, i.e., the zero vector space again, it implies that the vector space $\{0\}$ including all the successors of itself is the same $\{0\}$. With all that, the cardinality of $\{0\}$ is obviously 1.\bigskip

\noindent In this way, $\texttt{ST}_{\texttt{Hil}}$ can be considered to be the set of axioms $\texttt{ZF}$ wherein the axioms $\textbf{\texttt{Empty}}$ and $\textbf{\texttt{Inf}}$ have been replaced with their negations, $\neg\textbf{\texttt{Empty}}$ and $\neg\textbf{\texttt{Inf}}$. In symbols,\smallskip

\begin{equation} 
  \texttt{ST}_{\texttt{Hil}}
   =
   \texttt{ZF}_{\texttt{fin}}
   \mathrel{\mathop:}=
   \texttt{ZF}_{\texttt{base}}
   \cup
   \left\{
      \neg\textbf{\texttt{Empty}},
      \neg\textbf{\texttt{Inf}}
   \right\}
   \;\;\;\;  ,
\end{equation}
\smallskip

\noindent where $\texttt{ZF}_{\texttt{base}}$ denotes the ``basic'' set theory, namely,\smallskip

\begin{equation} 
  \texttt{ZF}_{\texttt{base}}
   \mathrel{\mathop:}=
   \texttt{ZF}
   \setminus
   \left\{
      \textbf{\texttt{Empty}},
      \textbf{\texttt{Inf}}
   \right\}
   \;\;\;\;  .
\end{equation}
\smallskip

\noindent While on the subject, let us note down that an $\omega$-model of $\texttt{ZF}_{\texttt{fin}}$ is a model in which every set has at most finitely many elements (as viewed externally).\bigskip

\noindent Consequently, the formal language of quantum mechanics, $\mathcal{L}(\texttt{QM})$, which has the axioms of Hilbert space theory at its core \cite{Edwards}, can be expressed by the formal language of the finite set theory $\texttt{ZF}_{\texttt{fin}}$ in conjunction with $\textbf{\texttt{X}}_{\texttt{qsystem}}$, the set of axioms of quantum mechanics that are absent in a set theory. This can be displayed by the following bijection:\smallskip

\begin{equation}  
   f\!\!:
   \mathcal{L}(\texttt{QM})
   \to
   \mathcal{L}
   \left(
      \texttt{ZF}_{\texttt{fin}}
      \cup
      \textbf{\texttt{X}}_{\texttt{qsystem}}
   \right)
   \;\;\;\;  .
\end{equation}
\smallskip

\noindent Consistent with the above mapping, the formal language of classical mechanics, $\mathcal{L}(\texttt{CM})$, must be expressed by the bijection\smallskip

\begin{equation}  
   f\!\!:
   \mathcal{L}(\texttt{CM})
   \to
   \mathcal{L}
   \left(
      \texttt{ZF}_{\texttt{fin}}
      \cup
      \textbf{\texttt{X}}_{\texttt{system}}
   \right)
   \;\;\;\;  ,
\end{equation}
\smallskip

\noindent meaning that everything which can be expressed through $\mathcal{L}(\texttt{CM})$ can also be expressed through the collection of axioms of the finite set theory $\texttt{ZF}_{\texttt{fin}}$ in conjunction with $\textbf{\texttt{X}}_{\texttt{system}}$, the set of axioms (non-existent in a set theory) determined by a classical mechanical system being a subject of study (the set $\textbf{\texttt{X}}_{\texttt{system}}$ is a proper subset of $\textbf{\texttt{X}}_{\texttt{qsystem}}$, i.e., $\textbf{\texttt{X}}_{\texttt{system}} \subsetneq\textbf{\texttt{X}}_{\texttt{qsystem}}$, otherwise classical mechanics cannot be reducible to quantum mechanics).\bigskip

\noindent In line with the reductionist approach \cite{Ney}, the transition from a classical field to a quantum operator field should be analogous to the promotion of a classical harmonic oscillator to a quantum harmonic oscillator. Accordingly, for each field theory $\texttt{FT}$ and its quantum counterpart $\texttt{QFT}$, the following bijections must hold\smallskip

\begin{equation}  
   f\!\!:
   \mathcal{L}(\texttt{FT})
   \to
   \mathcal{L}
   \left(
      \texttt{ZF}_{\texttt{fin}}
      \cup
      \textbf{\texttt{X}}_{\texttt{field}}
   \right)
   \;\;\;\;  ,
\end{equation}
\smallskip
\vspace{-26pt}

\begin{equation}  
   f\!\!:
   \mathcal{L}(\texttt{QFT})
   \to
   \mathcal{L}
   \left(
      \texttt{ZF}_{\texttt{fin}}
      \cup
      \textbf{\texttt{X}}_{\texttt{qfield}}
   \right)
   \;\;\;\;  ,
\end{equation}
\smallskip

\noindent where $\textbf{\texttt{X}}_{\texttt{field}}$ and $\textbf{\texttt{X}}_{\texttt{qfield}}$ denote the sets of axioms that are not present in a set theory, to be specific, $\textbf{\texttt{X}}_{\texttt{field}}$ comprises axioms that depend on a classical filed being studied and $\textbf{\texttt{X}}_{\texttt{qfield}}$ is the proper superset of $\textbf{\texttt{X}}_{\texttt{field}}$, i.e., $\textbf{\texttt{X}}_{\texttt{qfield}} \supsetneq \textbf{\texttt{X}}_{\texttt{field}}$.\bigskip

\noindent On condition that any $\texttt{FT}$ and $\texttt{QFT}$ are decidable, every well-formed formula in the formal languages $\mathcal{L}(\texttt{FT})$ and $\mathcal{L}(\texttt{QFT})$ must be consistent with the axiom $\neg\textbf{\texttt{Inf}}$. This implies that calculations made with $\texttt{FT}$ and $\texttt{QFT}$ are expected to be free from the infinite elements $+\infty$ and $-\infty$.\bigskip

\noindent To check this, let us calculate the zero-point energy $\langle 0|\hat{H}_{F}|0 \rangle$ of a particle field Hamiltonian $\hat{H}_{F}$. This calculation can be viewed as a summation over quantum harmonic oscillators with the zero-point energy $\hbar\omega/2$ at all points in space:\smallskip

\begin{equation} \label{CALC} 
   \langle 0|\hat{H}_{F}|0 \rangle
   =
   \frac{\hbar\omega}{2}
   \sum_{n=1} 
   1
   \;\;\;\;  .
\end{equation}
\smallskip

\noindent Since the axiom $\neg\textbf{\texttt{Inf}}$ is a part of quantum formalism $\mathcal{L}(\texttt{QFT})$, the existence of an endless sequence of units such as $1,1,1,\ldots = (1)_{n=1}^{\infty}$, where the symbol $\infty$ denotes an unbounded limit, is not allowed. The implication of this is that the series $\sum_{n=1}1$ must end up convergent. Thence, the vacuum energy ought to be finite. By contrast, if the axiom $\textbf{\texttt{Inf}}$, not its negation, were to be a part of $\mathcal{L}(\texttt{QFT})$, then the infinite sequence of additions $1+1+1+\dots = \sum_{n=1}^{\infty}1$ would have the right to be in (\ref{CALC}) resulting in the infinite zero-point energy.\bigskip

\noindent Then again, the zero-point energy could always be finite if space were to have a finite number of points, id est, if a geometry of physical space were to be finite.\bigskip

\noindent Before proceeding further, allow us to clarify the difference between finite geometries, bounded (finite) metric geometries and discrete geometries. For the purpose of the current presentation, one can define a geometry as a system of axioms that identify what ``things'' are which constitute ``points'', ``lines'', ``planes'', and so forth.\bigskip

\noindent In terms of this definition, a finite geometry is any of axiomatic systems that have only a finite number of points.\bigskip

\noindent A finite metric geometry, on the other hand, is an axiomatic system whose set of points is bounded, i.e., all of its points are within a certain (finite) distance of each other \cite{Burago}.\bigskip 

\noindent At the same time, a discrete geometry (including the causal set program \cite{Brightwell, Sorkin, Bezdek} whereby spacetime is a collection of points randomly selected in a background continuous space) takes up only objects in which points are isolated from each other in some sense, as for example, the set of natural numbers is a discrete set, i.e., a set of isolated points.\bigskip

\noindent Most important of all, neither a finite metric geometry nor a discrete geometry need to have a finite number of points.\bigskip

\noindent Unless otherwise stipulated, henceforth the statements will relate to finite and non-finite geometries alike.\bigskip

\noindent Let $M$ denote a manifold (such as a surface or a space). Assume that the manifold $M$ admits a notion of distance between its points; so, $M$ is equipped with measures of its regions (i.e., connected parts of $M$) such as area $A$ and volume $V$.\bigskip

\noindent Consider a manifold $\mathcal{R}$ that is taken to be a region in another manifold $M$, which means that $\mathcal{R}$ is deemed to be a subset of $M$ having the same dimension as $M$ does. For instance, $\mathcal{R}$ can be a 3-ball in Euclidean 3-space.\bigskip

\noindent Conceding that the characteristic length $L(\mathcal{R})$ defining the linear scale of the region $\mathcal{R}$ is the ratio of the region volume $V(\mathcal{R})$ to the area $A(\mathcal{R})$ of the region boundary, i.e.,\smallskip

\begin{equation}  
   L(\mathcal{R})
   =
   \frac{V(\mathcal{R})}{A(\mathcal{R})}
   \;\;\;\;  ,
\end{equation}
\smallskip

\noindent and on condition that $\omega$ is given by the expression\smallskip

\begin{equation}  
  \omega
   =
   \frac{2 \pi c}{L(\mathcal{R})}
   \;\;\;\;  ,
\end{equation}
\smallskip

\noindent one finds\smallskip

\begin{equation}  
   \langle 0|\hat{H}_{F}|0 \rangle
   =
   \frac{\pi\hbar c}{L(\mathcal{R})}
   \cdot
   P_{\mathrm{vac}}
   \;\;\;\;  .
\end{equation}
\smallskip

\noindent where $P_{\mathrm{vac}}$ denotes the series $1+1+\dots =\sum_{n=1}1$.\bigskip

\section{The holographic principle}  

\noindent Naturally, the zero-point energy $\langle 0|\hat{H}_{F}|0 \rangle$ can be presented as the result of multiplying the vacuum energy density $\rho_{\mathrm{vac}}$ by the region volume $V(\mathcal{R})$. So, by allowing $\rho_{\mathrm{vac}}$ to be proportional to the effective cosmological constant $\Lambda_{\mathrm{eff}}$, namely,\smallskip

\begin{equation}  
   \rho_{\mathrm{vac}}
   =
   \frac{c^{4}}{8\pi G}
   \cdot
   \Lambda_{\mathrm{eff}}
   \;\;\;\;  ,
\end{equation}
\smallskip

\noindent one can express \emph{the cardinality of the region $\mathcal{R}$} (the number of points constituting $\mathcal{R}$) in terms of $\Lambda_{\mathrm{eff}}$:\smallskip

\begin{equation} \label{CARD} 
   P_{\mathrm{vac}}
   =
   \frac{\Lambda_{\mathrm{eff}} \cdot V(\mathcal{R}) \cdot L(\mathcal{R})}{8 \pi^2 \ell_P^2}
   \;\;\;\;  ,
\end{equation}
\smallskip

\noindent where $\ell_P = \sqrt{\hbar G/c^3}$.\bigskip

\noindent At this juncture, let us establish two positive dimensionless ratios:\smallskip

\begin{equation}  
   \alpha
   \mathrel{\mathop:}=
   \frac{\Lambda_{\mathrm{eff}} \cdot D_U^2}{8\pi^2}
   \;\;\;\;  ,
\end{equation}
\vspace{-13pt}

\begin{equation}  
   \xi(\mathcal{R})
   \mathrel{\mathop:}=
   \frac{L(\mathcal{R})}{D_U}
   \;\;\;\;  ,
\end{equation}
\smallskip

\noindent in which $D_U$ stands for the comoving (``static'') diameter of the observable universe. As $\Lambda_{\mathrm{eff}}$ and $D_U$ are both nonnegative constants, the ratio $\alpha$ is expected to remain constant in time. Using $\alpha$ and $\xi(\mathcal{R})$, one can present (\ref{CARD}) as the equality\smallskip

\begin{equation} \label{HOL} 
   P_{\mathrm{vac}}
   =
   \rho_{V}^{\text{ }}(\mathcal{R}) \cdot V(\mathcal{R})
   =
   \rho_{\!A}^{\text{ }}(\mathcal{R}) \cdot A(\mathcal{R})
   \;\;\;\;  ,
\end{equation}
\smallskip

\noindent where $\rho_V(\mathcal{R})$ and $\rho_A(\mathcal{R})$ denote the volumetric density and the areal density of the cardinality of the region $\mathcal{R}$, respectively,

\begin{equation} \label{DV} 
   \rho_{V}^{\text{ }}(\mathcal{R})
   =
   \alpha
   \cdot
   \frac{\xi^2(\mathcal{R})}{L(\mathcal{R}) \cdot \ell_P^2}
   \;\;\;\;  ,
\end{equation}
\vspace{-13pt}

\begin{equation} \label{DA} 
   \rho_{\!A}^{\text{ }}(\mathcal{R})
   =
   \alpha
   \cdot
   \frac{\xi^2(\mathcal{R})}{\ell_P^2}
   \;\;\;\;  .
\end{equation}
\smallskip

\noindent This enables the holographic principle: points that compose a region are entirely contained in the boundary to the region.\bigskip

\section{The entropy in a manifold}  

\noindent Let $\mathcal{C}(\mathcal{R})$ be a system of coordinates, i.e., a set of numbers, that specify the position of each point in a region $\mathcal{R}$. To be qualified as systems of coordinates, $\mathcal{C}(\mathcal{R})$ must be such that the operations of addition, subtraction, multiplication, and division are defined and satisfy the closure under addition and subtraction. Considering this, the system of coordinates $\mathcal{C}(\mathcal{R})$ must be a field \cite{Mullen}.\bigskip

\noindent In a geometry that has only a finite number of points, the cardinality of the field $\mathcal{C}(\mathcal{R})$ is a prime power, i.e.,\smallskip

\begin{equation}  
   \mathrm{card}\left( \mathcal{C}(\mathcal{R}) \right)
   =
   p^q
   \;\;\;\;  ,
\end{equation}
\smallskip

\noindent where $p$ is a prime number and $q$ is the number of points comprising the region $\mathcal{R}$. This means that the system of coordinates $\mathcal{C}(\mathcal{R})$ has $p^q$ elements, $q\log{p}$ degrees of freedom and can store $q\log_2{p}$ bits of information (that is, $\log_2{p}$ bits per each point in $\mathcal{R}$).\bigskip

\noindent A measure of ignorance about the position of the points in $\mathcal{R}$ can be taken to be proportional to $2^q$, i.e., the smallest possible size of a coordinate system for a given region. For that reason, the entropy in the region $\mathcal{R}$ can be determined as\smallskip

\begin{equation} \label{ENTM} 
   H(\mathcal{R})
   =
   k_B
   \cdot
   \log_2{\mathrm{card}\left( \mathcal{C}(\mathcal{R}) \right)}
   =
   k_B
   \cdot
   q
   \;\;\;\;  ,
\end{equation}
\smallskip

\noindent where $k_B$ is the Boltzmann constant. The above allows to construe each point in space as a bit of information.\bigskip

\noindent Recalling that $q = P_{\mathrm{vac}}$ and using the equality (\ref{HOL}), one finds that the entropy in a volume $V(\mathcal{R})$ is equivalent to the entropy in the area $A(\mathcal{R})$ of the boundary to $\mathcal{R}$. In symbols,\smallskip

\begin{equation}  
   k_B
   \cdot
   P_{\mathrm{vac}}   
   =
   H(V(\mathcal{R}))
   =
   H(A(\mathcal{R}))
   \;\;\;\;  ,
\end{equation}
\smallskip

\noindent where $H(V(\mathcal{R}))$ and $H(A(\mathcal{R}))$ denote the entropy in $V(\mathcal{R})$ and $A(\mathcal{R})$ such that\smallskip

\begin{equation}  
   H(V(\mathcal{R}))
   =
   k_B
   \cdot
   \rho_{V}^{\text{ }} V(\mathcal{R})
   =
   k_B
   \cdot
   \frac{\alpha\xi^2(\mathcal{R})}{L(\mathcal{R}) \cdot \ell_P^2}
   \cdot
   V(\mathcal{R})
   \;\;\;\;  ,
\end{equation}
\vspace{-13pt}

\begin{equation} \label{ENTRA} 
   H(A(\mathcal{R}))
   =
   k_B
   \cdot
   \rho_{\!A}^{\text{ }} A(\mathcal{R})
   =
   k_B
   \cdot
   \frac{\alpha\xi^2(\mathcal{R})}{\ell_P^2}
   \cdot
   A(\mathcal{R})
   \;\;\;\;  .
\end{equation}
\smallskip

\section{The cosmological constant problem}  

\noindent Thanks to the factor $\kappa = 8\pi G /c^4$, the upper bound on the effective cosmological constant $\Lambda_{\mathrm{eff}} = \kappa\cdot\rho_{\mathrm{vac}}$ laid by observations is interpreted as an observational bound on the vacuum energy density $\rho_{\mathrm{vac}}$. The problem is that the zero-point energy calculated along the lines of the axiom of infinity $\textbf{\texttt{Inf}}$ comes out infinite: Recall that the series $P_{\mathrm{vac}} \mathrel{\mathop:}= 1+1+\dots =\sum_{n=1}1$ diverges under the assumption that $\textbf{\texttt{Inf}}$ holds true. And even though the formalism based on $\textbf{\texttt{Inf}}$ can produce (at the quantitative level) finite values of $\rho_{\mathrm{vac}}$ using one or another renormalization scheme, such finite, renormalized, values end up being greater than the observational bound by at least 40 orders of magnitude \cite{Rugh, Carroll, Peebles}. This constitutes the conundrum known as ``the cosmological constant problem''.\bigskip

\noindent In contrast, within the formalism wherein the axiom $\neg\textbf{\texttt{Inf}}$ is a part, the number of points $P_{\mathrm{vac}}$ constituting space is finite, which, in turn, indicates that the vacuum energy is intrinsically finite.\bigskip

\noindent What is more, on large scales, the space wherein the universe lives is well approximated as three-dimensional and flat \cite{Liddle}. Given that, the manifold $M_U$ associated with the universe can be estimated as a 3-dimensional Euclidean space. By the same token, the region $\mathcal{R}_U \subset M_U$ representing the space of the observable universe can be looked upon as an ordinary ball (i.e., a 3-ball) of volume $V(\mathcal{R}_U)$ with the boundary of area $A(\mathcal{R}_U)$. Accordingly, $L(\mathcal{R}_U)=R_U/3$ and $\xi(\mathcal{R}_U) = 1/6$.\bigskip

\noindent Hence, considering that (a) a black hole is the most entropic object one can put inside the spherical surface enclosing the region of the observable universe $\mathcal{R}_U$, and (b) the observable universe is not a black hole, one can find the upper bound of the entropy in the area $A(\mathcal{R}_U)$:\smallskip

\begin{equation} \label{BOUND} 
   H(A(\mathcal{R}_U))
   <
   H_{\mathrm{BH}}(A(\mathcal{R}_U))
   \;\;\;\;  ,
\end{equation}
\smallskip

\noindent where $H_{\mathrm{BH}}$ is the Bekenstein-Hawking entropy \cite{Bekenstein, Hawking}\smallskip

\begin{equation} \label{BHE} 
   H_{\mathrm{BH}}(A(\mathcal{R}))
   =
   k_B
   \cdot
   \frac{A(\mathcal{R})}{4\ell_P^2}
   \;\;\;\;   
\end{equation}
\smallskip

\noindent in which $\mathcal{R} = \mathcal{R}_U$\bigskip.

\noindent As to the entropy $H(A(\mathcal{R}_U))$, using (\ref{ENTRA}) it can be evaluated as\smallskip

\begin{equation}  
   H(A(\mathcal{R}_U))
   =
   k_B
   \cdot
   \alpha
   \cdot
   \frac{1}{6^2\ell_P^2}
   \cdot
   A(\mathcal{R}_U)
   \;\;\;\;  .
\end{equation}
\smallskip

\noindent Thus, the bound (\ref{BOUND}) produces\smallskip

\begin{equation}  
   \alpha
   <
   9
   \;\;\;\;  .
\end{equation}
\smallskip

\noindent Due to the fact that the ratio $\alpha$ is on a par with 1, the effective cosmological constant $\Lambda_{\mathrm{eff}}$ is close to zero. To wit,\smallskip

\begin{equation}  
   \Lambda_{\mathrm{eff}}
   =
   \frac{8\pi^2}{D_{U}^{2}}
   \cdot
   \alpha
   <
   \frac{18\pi^2}{R_{U}^{2}}
   \sim
   9.3 \times 10^{-52}\, \mathrm{m}^{-2}
   \;\;\;\;  ,
\end{equation}
\smallskip

\noindent where $R_U$ is the comoving radius of the observable universe.\bigskip

\noindent The above result exceeds the bound implied by cosmological observations $\Lambda_{\mathrm{eff}} \apprle 10^{-52}\, \mathrm{m}^{-2}$ \cite{Planck} by only one order of magnitude.\bigskip

\section{The volume inside a black hole}  

\noindent The relation between geometry and information may help to define a meaningful notion of volume inside a black hole.\bigskip

\noindent For convenience of reference, let us denote the region of a black hole by $\mathcal{R}_{\mathrm{BH}}$. Unlike $A(\mathcal{R}_{\mathrm{BH}})$, the area of the horizon (i.e., surface) of a black hole remaining the same for all observers, the volume inside a black hole, $V(\mathcal{R}_{\mathrm{BH}})$, is not a precisely defined concept: $V(\mathcal{R}_{\mathrm{BH}})$ depends on an arbitrary choice of coordinates, and as such can be time dependent, or even zero \cite{Parikh, Dinunno, Christodoulou}.\bigskip

\noindent This implies that the characteristic length of the black hole's interior,\smallskip

\begin{equation}  
   L(\mathcal{R}_{\mathrm{BH}})
   =
   \frac{V(\mathcal{R}_{\mathrm{BH}})}{A(\mathcal{R}_{\mathrm{BH}})}
   \;\;\;\;  ,
\end{equation}
\smallskip

\noindent is undefined and so is the ratio $\xi(\mathcal{R}_{\mathrm{BH}}) = L(\mathcal{R}_{\mathrm{BH}}) / 2R_U$. As a result, one can only argue that the amount of information inside a black hole is contained on the surface of the black hole, namely,\smallskip

\begin{equation}  
   H(V(\mathcal{R}_{\mathrm{BH}}))
   =
   H(A(\mathcal{R}_{\mathrm{BH}}))
   \;\;\;\;  .
\end{equation}
\smallskip

\noindent This begs the question: If $V(\mathcal{R}_{\mathrm{BH}})$ were to be undefined, how could this expression be true?\bigskip

\noindent But then, in accordance with (\ref{ENTRA}) the said amount of information can be calculated as\smallskip

\begin{equation}  
   H(A(\mathcal{R}_{\mathrm{BH}}))
   =
   k_B
   \cdot
   \alpha
   \cdot
   \frac{L^2(\mathcal{R}_{\mathrm{BH}})}{4R_U^2}
   \cdot
   \frac{A(\mathcal{R}_{\mathrm{BH}})}{\ell_P^2}
    \;\;\;\;  .
\end{equation}
\smallskip

\noindent Comparing the above with the Bekenstein-Hawking entropy (\ref{BHE}) in which $\mathcal{R}=\mathcal{R}_{\mathrm{BH}}$ gives\smallskip

\begin{equation}  
   L(\mathcal{R}_{\mathrm{BH}})
   =
   \frac{R_U}{\sqrt{\alpha}}
   >
   \frac{R_U}{3}   
    \;\;\;\;  .
\end{equation}
\smallskip

\noindent Hence, the Bekenstein-Hawking assumption of black hole entropy is equivalent to the supposition that $L(\mathcal{R}_{\mathrm{BH}})$ is one and the same for all black holes and it is commensurate with the comoving radius of the observable universe.\bigskip

\noindent Therefore, the unique interior volume of a black hole (invariable for all observers) can be believed to be\smallskip

\begin{equation} \label{VBH} 
   V(\mathcal{R}_{\mathrm{BH}})
   =
   \frac{R_U \cdot A(\mathcal{R}_{\mathrm{BH}})}{\sqrt{\alpha}}
   >
   \frac{1}{3}
   R_U \cdot A(\mathcal{R}_{\mathrm{BH}})
   \;\;\;\;  .
\end{equation}
\smallskip

\noindent Owing to the term $h > R_U/3$, the volume $V(\mathcal{R}_{\mathrm{BH}}) = h \cdot A(\mathcal{R}_{\mathrm{BH}})$ is extremely large. By way of example, for a Schwarzschild black hole with the area of the horizon of radius $R_S \approx 3\, \mathrm{km}$, $V(\mathcal{R}_{\mathrm{BH}})$ is greater than the volume of the sphere in 3-space $\mathbb{R}^3$ with the radius $ r_{\text{sphere}}$ comparable with the distance from the Sun to Mars:\smallskip

\begin{equation}  
    r_{\text{sphere}}
   >
   \sqrt[3]
   {R_U R_S^2}
   \approx
   226\,\text{million km}
   \;\;\;\;  .
\end{equation}
\smallskip

\noindent Concerning a geometric shape of the interior region of a black hole, it can be imagined in agreement with the formula (\ref{VBH}) as a figure resembling a cylinder with the base of area $A(\mathcal{R}_{\mathrm{BH}})$ and the height $h$. Providing the surface area of the hole's interior coincides with the area of the event horizon, the surface area of the cylinder should be made up of just one component, $A(\mathcal{R}_{\mathrm{BH}})$. This means that the said cylinder must be devoid of its side and one of its bases. Needless to say, a geometric shape like this cannot exist in a three-dimensional Euclidean space.\bigskip

\section{Magnification of a holographic image}  

\noindent The product of $\rho_V(\mathcal{R})$ and $V(\mathcal{R})$ determines the amount of information contained in a region $\mathcal{R}$ of volume $V$. Therefore, $\rho_V(\mathcal{R})$ can be considered to be the three-dimensional (volumetric) density of information. Likewise, $\rho_A(\mathcal{R})$ can be seen as the two-dimensional (areal) density of information contained in the boundary to the region $\mathcal{R}$.\bigskip

\noindent Consider $\mathcal{R}=\mathcal{R}_U$. In line with (\ref{DV}) and (\ref{DA}), the degrees of information concentration in $\mathcal{R}_U$ are:\smallskip

\begin{equation}  
   \rho_{V}^{\text{ }}(\mathcal{R}_U)
   =
   \alpha
   \cdot
   \frac{1}{12R_U\ell_P^2}
   \;\;\;\;  ,
\end{equation}
\vspace{-13pt}

\begin{equation}  
   \rho_{\!A}^{\text{ }}(\mathcal{R}_U)
   =
   \alpha
   \cdot
   \frac{1}{6^2\ell_P^2}
   \;\;\;\;  .
\end{equation}
\smallskip

\noindent From here it follows that the area containing 1 bit of information is\smallskip

\begin{equation}  
   A_{1}
   =
   \alpha^{-1}\left( 6 \ell_P \right)^2
   \;\;\;\;  ,
\end{equation}
\smallskip

\noindent at the same time as the volume composed of 1 bit of information is\smallskip

\begin{equation}  
   V_{1}
   =
   \frac{1}{3}
   R_U
   \cdot
   \alpha^{-1}\left( 6 \ell_P \right)^2
   \;\;\;\;  .
\end{equation}
\smallskip

\noindent In a mathematical formalism embracing the axiom $\neg\textbf{\texttt{Empty}}$, the minimum cardinality of a manifold must be 1. This requires that all $A$ and $V$ must be bounded from below such that $A\ge A_1$ and $V\ge V_1$. Providing sphericalness of $A$ and $A_1$, as well as $V$ and $V_1$, it means $r_{\!A}^{\text{ }}\ge r_{\!A_{1}}^{\text{ }}$ and $r_{\!V}^{\text{ }}\ge r_{\!V_{1}}^{\text{ }}$, where\smallskip

\begin{equation}  
   r_{\!A_{1}}^{\text{ }}
   =
   \frac{3}{\sqrt{\pi\alpha}}\ell_P
   >
   \frac{1}{\sqrt{\pi}}\ell_P
   \approx
   \ell_P
   \;\;\;\;  ,
\end{equation}
\vspace{-7pt}

\begin{equation}  
   r_{\!V_{1}}^{\text{ }}
   =
   \sqrt[3]
   {\frac{1}{4\pi\alpha}R_U\cdot\left( 6 \ell_P \right)^2} 
   >
   \sqrt[3]
   {\frac{1}{\pi}R_U \cdot \ell_P^2}
   \sim
   3.3\times10^{-15}\,\text{m}
   \approx
   \ell_s
   \;\;\;\;  .
\end{equation}
\smallskip

\noindent As appears, the minimal length scale in a two-dimensional manifold is the Planck length $\ell_P$, while the said scale in a three-dimensional manifold is equivalent to the approximate limit of the strong interaction $\ell_s\sim 3.0\times 10^{-15}$ m. That is to say, there are two different minimal length scales in physical space: $\ell_P$ and $\ell_s$. Such cannot be unless the sphere of radius $r_{\!V_{1}}^{\text{ }}\apprge\ell_s$ is an enlarged holographic image of the circle of radius $r_{\!A_{1}}^{\text{ }}\apprge\ell_P$.\bigskip

\noindent This enlargement explains the puzzlement of the holographic principle. Definitely, suppose, for a moment, that there is no enlargement. Then, covering a surface of the observable universe with primitive circles of radius $r_{\!A_{1}}^{\text{ }}$ would be holographically projected as filling in the interior of the universe with spheres of radius $r_{\!A_{1}}^{\text{ }}$. Since $r_{\!A_{1}}^{\text{ }} \ll R_U$, the universe's interior could be considered as a set of points with the cardinality greatly surpassing the number of points that constitute the universe's surface:\smallskip

 \begin{equation}  
   \frac{R_U^3}{r_{\!A_{1}}^{3}}
   =
   \frac{4\pi R_U^2}{\alpha^{-1}\left( 6 \ell_P \right)^2}
   \cdot
   \frac{R_U}{r_{\!A_{1}}^{\text{ }}}
   \gg
   \frac{4\pi R_U^2}{A_{1}}
   \;\;\;\;  .
\end{equation}
\smallskip

\noindent Obviously, that would be contradictory to the holographic principle. However, due to the holographic enlargement, the universe's interior is discretized by spheres of radius $r_{\!V_{1}}^{\text{ }}$, therefore,\smallskip

 \begin{equation}  
   \frac{R_U^3}{r_{\!V_{1}}^{3}}
   =
   \frac{4\pi R_U^2}{\alpha^{-1}\left( 6 \ell_P \right)^2}
   \cdot
   \frac{R_U}{R_U}
   =
   \frac{4\pi R_U^2}{A_{1}}
   \;\;\;\;  .
\end{equation}
\smallskip

\noindent Magnification of a holographic image projected from the surface of the observable universe into its interior can be quantified by the ratio\smallskip

\begin{equation}  
   \frac{r_{\!V}^{\text{ }}}{r_{\!A}^{\text{ }}}
   =
   \sqrt[3]
   {\frac{R_U}{r_{\!A}}}
   \;\;\;\;  .
\end{equation}
\smallskip

\noindent The above means that if the size of the image is $r_{\!V}$, then the ``true'' size of a projected object $r_{\!A}$ can be estimated as\smallskip

\begin{equation} \label{TRUES}  
   r_{\!A}^{\text{ }}
   =
   \sqrt
   {\frac{r_{\!V}^3}{R_U}}
   \;\;\;\;  .
\end{equation}
\smallskip

\noindent For example, the Solar System whose radius is $r_{\!V}^{\text{ }} \sim 4.5$ billion kilometers can be believed to be a holographic image of a two-dimension structure of size $r_{\!A}^{\text{ }} \sim 455.1$ kilometers ``painted'' on the boundary to the observable universe.\bigskip

\noindent By the same token, the ``true'' distance between the Sun and the nearest known planetary system (Proxima Centauri system) equal to 4.2 light-years when measured through a region of space would be 377.6 million kilometers (about the distance from the Sun to the asteroid belt occupying the orbit between Mars and Jupiter) if it were to be measured on the surface of the observable universe.\bigskip

\noindent In passing, it should be noted this. Because the weak force has an effective range $\ell_w$ which is shorter than $\ell_s$ (namely, $\ell_w$ is around $10^{-17}$ to $10^{-16}$ m), it can be inferred that the weak interaction takes place only on a surface of the observable universe. Certainly, had $\ell_w^3$ been a holographic projection of some area on the surface of the observable universe, its ``true'' scale would have been $\sim 1.5 \times 10^{-39}$ m in accordance with (\ref{TRUES}), i.e., much less than the Planck length $\ell_P$. This may explain why the weak interaction does not produce a bound state, i.e., one in which a particle tends to remain localized in a region of space.\bigskip

\section{ Concluding remarks}  

\noindent The following critical comments could be passed on the claims put forward in this paper.\bigskip

\noindent The existence of a minimal length scale modifies the Heisenberg uncertainty principle, so that it is impossible to localize a test particle in essence. Subsequently, the notion of ``point'' breaks down causing the cardinality of a physical manifold (i.e., the number of points constituting this manifold) to become an ill-defined measure. Due to this, talking about the finiteness of the physical manifold has little sense.\bigskip

\noindent To reply to this criticism, let us first recall that gravity -- instead of spoiling the renormalizability of quantum field theories -- has long been suggested to lead to an effective cutoff in the ultraviolet, i.e. a minimal length scale $\ell_0$ \cite{Kempf, Nozari}. Logically $\ell_0$ implies a nonzero minimal uncertainty $\Delta x_0$ in position measurements. The latter can be implemented by generalizing the Heisenberg uncertainty principle.\bigskip

\noindent For example, in one dimension, the simplest generalized uncertainty relation incorporates a new term in the right hand side proportional to $(\Delta p)^2$, namely, $\Delta x \Delta p \ge \hbar/2 + \beta (\Delta p)^2$, where $\beta$ is a positive parameter independent of $\Delta x$ and $\Delta p$. So, when $\Delta p$ increases, the new term precludes $\Delta x$ from becoming arbitrarily small. This results in a nonzero minimal uncertainty $\Delta x_0$.\bigskip

\noindent Be that as it may, the existence of $\Delta x_0$ does not necessarily mean that there is a minimal length scale $\ell_0$ \cite{Hossenfelder}. In fact, the implication $\ell_0 \to \Delta x_0$ is identical to that of $\neg \ell_0 \vee \Delta x_0$. The last reads: ``It is true to say that there is $\Delta x_0$ but no such thing as $\ell_0$''.\bigskip

\noindent Secondly, in accordance with the expression for the entropy in the manifold proposed in this paper (see for example Eq. (\ref{ENTM})), each point constituting the physical manifold equates with a bit of information. Accordingly, the finiteness of the manifold is the notion defined as well as it gets. By way of illustration, the finiteness of the space of the observable universe signifies that the amount of information required to describe this space is limited.\bigskip

\noindent Thirdly, the existence of cutoffs is necessary for treating infinities that arise in calculated quantities of quantum field theories. Whatever such cutoffs are, they all involve nontrivial assumptions like the presence of unknown new physics \cite{Veltman, Cao}. Meanwhile, another, almost trivial way to avoid infinities in physics in the first place is to negate the axiom of infinity $\textbf{\texttt{Inf}}$ of Zermelo-Fraenkel set theory. This is exactly what has been demonstrated in the present paper.\bigskip

\section*{Conflict of interest}

\noindent The author states that there is no conflict of interest.\bigskip

\section*{Data Availability Statement}

\noindent The original contributions presented in the study are included in the article, further inquiries can be directed to the corresponding author.\bigskip

\bibliographystyle{References}
\bibliography{Hollographic_principle_Major_revision}

\begin{thebibliography}{10}
\expandafter\ifx\csname urlstyle\endcsname\relax
  \providecommand{\doi}[1]{doi:\discretionary{}{}{}#1}\else
  \providecommand{\doi}{doi:\discretionary{}{}{}\begingroup
  \urlstyle{rm}\Url}\fi

\bibitem{Hooft}
Gerard {'t Hooft}.
\newblock {Black holes and the dimensionality of space-time}.
\newblock In Ulf Lindstr{\"o}m, editor, \emph{{The Oskar Klein Centenary.
  Proceedings of the Symposium, 19-21 Sept. 1994, Stockholm, Sweden}}, pages
  122--137. World Scientific, 1995.

\bibitem{Susskind}
Leonard Susskind.
\newblock {The World as a Hologram}.
\newblock \emph{{J. Math. Phys.}}, 36:6377--6396, 1995.

\bibitem{Bigatti}
Daniela Bigatti and Leonard Susskind.
\newblock {TASI lectures on the Holographic Principle}.
\newblock {https://arxiv.org/abs/hep-th/0002044}, Feb 2000.

\bibitem{Bousso}
Raphael Bousso.
\newblock {The holographic principle}.
\newblock \emph{{Reviews of Modern Physics}}, 74:825--874, 2002.

\bibitem{Kaku}
Michio Kaku.
\newblock \emph{{Introduction to Superstrings and M-Theory}}.
\newblock {Springer}, 2011.

\bibitem{Ammon}
Martin Ammon and Johanna Erdmenger.
\newblock \emph{{Gauge/Gravity Duality: Foundations and Applications (1 ed.)}}.
\newblock {Cambridge University Press}, 2015.

\bibitem{Maldacena}
Juan Maldacena.
\newblock {The Large N limit of superconformal field theories and
  supergravity}.
\newblock \emph{{Advances in Theoretical and Mathematical Physics}},
  2:231--252, 1998.

\bibitem{McLerran}
Larry McLerran.
\newblock {Theory Summary: Quark Matter 2006}.
\newblock \emph{Journal of Physics G: Nuclear and Particle Physics},
  34(8):S583--S592, 2007.

\bibitem{Bolotin}
Arkady Bolotin.
\newblock {Physics in a finite geometry}.
\newblock {https://arxiv.org/abs/2212.02915}, Dec 2022.

\bibitem{Weisstein}
Eric~W. Weisstein.
\newblock {Urelement}.
\newblock In \emph{{From MathWorld -- A Wolfram Web Resource}}.
  {https://mathworld.wolfram.com/Urelement.html}, 2022.

\bibitem{Taylor}
Ralph~Gregory Taylor.
\newblock {Zermelo, Reductionism, and the Philosophy of Mathematics}.
\newblock \emph{{Notre Dame Journal of Formal Logic}}, 34(4):539--563, 1993.

\bibitem{Edwards}
David Edwards.
\newblock {The Mathematical Foundations of Quantum Mechanics}.
\newblock \emph{Synthese}, 42(1):1--70, 1979.

\bibitem{Ney}
Alyssa Ney.
\newblock {Reductionism}.
\newblock In \emph{{The Internet Encyclopedia of Philosophy. ISSN 2161-0002}}.
  {https://iep.utm.edu/red-ism/}, 2022.

\bibitem{Burago}
Dmitri Burago, Yuri Burago, and Sergei Ivanov.
\newblock \emph{{A Course in Metric Geometry}}.
\newblock {American Mathematical Society}, {Providence, Rhode Island}, 2000.

\bibitem{Brightwell}
Graham Brightwell and Ruth Gregory.
\newblock {Structure of Random Discrete Spacetime}.
\newblock \emph{Physical Review Letters}, 66(3):260--263, 1991.

\bibitem{Sorkin}
Raphael Sorkin.
\newblock {Causal sets: Discrete gravity}.
\newblock In A.~Gomberoff and D.Marolf, editors, \emph{{Lectures on Quantum
  Gravity}}, volume {Pan-American Advanced Studies Institute, School on Quantum
  Gravity}, pages 305--327. {Springer}, 2005.

\bibitem{Bezdek}
K{\'{a}}roly Bezdek.
\newblock \emph{{Classical Topics in Discrete Geometry}}.
\newblock {Springer}, New York, NY, 2010.

\bibitem{Mullen}
Gary~L. Mullen and Daniel Panario.
\newblock \emph{{Handbook of Finite Fields}}.
\newblock {Chapman and Hall/CRC}, New York, 2013.

\bibitem{Rugh}
Svend~E. Rugh and Henrik Zinkernagel.
\newblock {The Quantum Vacuum and the Cosmological Constant Problem}.
\newblock \emph{{Studies in History and Philosophy of Modern Physics}},
  33(4):663--705, 2001.

\bibitem{Carroll}
Sean~M. Carroll.
\newblock {The Cosmological Constant}.
\newblock \emph{{Living Reviews in Relativity}}, 4(1):1--56, 2001.

\bibitem{Peebles}
Phillip James~Edwin Peebles and Bharat Ratra.
\newblock {The cosmological constant and dark energy}.
\newblock \emph{{Reviews of Modern Physics}}, 75(2):559--606, 2003.

\bibitem{Liddle}
Andrew Liddle.
\newblock \emph{{An Introduction to Modern Cosmology. Third Edition}}.
\newblock {Wiley}, UK, 2015.

\bibitem{Bekenstein}
Jacob~D. Bekenstein.
\newblock {Holographic bound from second law of thermodynamics}.
\newblock \emph{{Physics Letters B}}, 481:339--345, 2000.

\bibitem{Hawking}
Stephen~W. Hawking.
\newblock {Black hole explosions?}
\newblock \emph{{Nature}}, 248(5443):30--31, 1974.

\bibitem{Planck}
{Planck Collaboration: P. A. R. Ade et al}.
\newblock {Planck 2015 results - XIII. Cosmological parameters}.
\newblock \emph{{Astronomy \& Astrophysics}}, 594:A13, 2016.

\bibitem{Parikh}
Maulik~K. Parikh.
\newblock {Volume of black holes}.
\newblock \emph{{Physical Review D}}, 73(124021):1--5, 2006.

\bibitem{Dinunno}
Brandon DiNunno and Richard~A. Matzner.
\newblock {The Volume Inside a Black Hole}.
\newblock \emph{{General Relativity and Gravitation}}, 42:63--76, 2010.

\bibitem{Christodoulou}
Marios Christodoulou and Carlo Rovelli.
\newblock {How big is a black hole?}
\newblock \emph{{Phys. Rev. D}}, 91(064046):1--7, 2015.

\bibitem{Kempf}
Achim Kempf, Gianpiero Mangano, and Robert~B. Mann.
\newblock {Hilbert Space Representation of the Minimal Length Uncertainty
  Relation}.
\newblock \emph{{Phys. Rev. D}}, 52(1108), 1995.

\bibitem{Nozari}
Kourosh Nozari and Amir Etemadi.
\newblock {Minimal length, maximal momentum and Hilbert space representation of
  quantum mechanics}.
\newblock \emph{{Phys. Rev. D}}, 85(104029), 2012.

\bibitem{Hossenfelder}
Sabine Hossenfelder.
\newblock {Minimal Length Scale Scenarios for Quantum Gravity}.
\newblock \emph{Living Reviews in Relativity}, 16(2):1--90, 2013.

\bibitem{Veltman}
G.~{'t Hooft} and M.~Veltman.
\newblock {Regularization and renormalization of gauge fields}.
\newblock \emph{{Nuclear Physics B}}, 44:189--213, 1972.

\bibitem{Cao}
Tian~Yu Cao and Silvan~S. Schweber.
\newblock {The Conceptual Foundations and the Philosophical Aspects of
  Renormalization Theory}.
\newblock \emph{{Synthese}}, 97:33--108, 1993.

\end{thebibliography}

\end{document}